\documentclass[12pt]{article}
\usepackage{graphics}
\usepackage{amssymb}
\usepackage{psfrag}

\newcommand{\beq}{\begin{equation}}
\newcommand{\eeq}{\end{equation}}
\newcommand{\bea}{\begin{eqnarray}}
\newcommand{\eea}{\end{eqnarray}}

\begin{document}

{\normalsize \hfill SPIN-06/10}\\
\vspace{-1.5cm}
{\normalsize \hfill ITP-UU-06/12}\\
${}$\\

\begin{center}
\vspace{60pt}
{ \Large \bf Unexpected Spin-Off from Quantum Gravity}

\vspace{60pt}

{\sl D. Benedetti}$\,^{a}$
and {\sl R. Loll}$\,^{a}$

\vspace{24pt}
{\footnotesize

$^a$~Institute for Theoretical Physics, Utrecht University, \\
Leuvenlaan 4, NL-3584 CE Utrecht, The Netherlands.\\
{email: d.benedetti@phys.uu.nl, r.loll@phys.uu.nl}\\

\vspace{10pt}

}
\vspace{58pt}

\end{center}

\begin{center}
{\bf Abstract}
\end{center}
\noindent
We propose a novel way of investigating the universal properties
of spin systems by coupling them to an ensemble of causal
dynamically triangulated lattices, instead of studying them on a fixed regular
or random lattice. Somewhat surprisingly,
graph-counting methods to extract high- or low-temperature
series expansions can be adapted to this case.
For the two-dimensional Ising model, we present evidence
that this ameliorates the singularity structure of thermodynamic functions
in the complex plane, and improves the convergence
of the power series.

\vspace{12pt}
\noindent

%\vfill

\newpage

\subsection*{Introduction}

Quantum gravity is not usually regarded a subject of particular
relevance to physics outside the exotic realm of Planck scale
phenomena. Nevertheless, progress in science can sometimes
come from an unexpected direction. In this letter, we will give
a concrete example of how a specific way of constructing a
theory of quantum gravity may lead to a new method for
understanding the critical behaviour of certain spin and matter
systems.

Apart from a handful of well-known exceptions, most thermodynamic
properties of statistical mechanical models are not known
to us in exact, closed form. Consequently, we must rely on a
variety of approximation and numerical methods to study their
behaviour. Lattice structures appearing in such models can either
reflect the actual microscopic composition of a particular magnetic
material, say, or play the role of a convenient discrete regulator
for a continuous system whose symmetry properties have
little to do with those of the lattice approximation.

Our focus will be on instances of the latter, where
one is only interested in universal
properties of the lattice models, which pertain to the system in
a suitable scaling limit and are largely independent of
discretization details, including those of the geometry of the
underlying lattice. This is a point of view also
encountered in the nonperturbative lattice formulations
of both quantum gauge theories and gravity in high-energy physics.

In such a ``utilitarian" view of lattice systems one is naturally led
to ask how one should set up the lattice discretization to
extract the desired continuum information in the quickest and
most reliable way. Part of this quest is a systematic study of
the influence of the lattice geometry on results, in an effort to
separate as cleanly as possible `universal behaviour' from
`lattice artefacts'. An example are the investigations of
two-dimensional Ising models \cite{ising} on different
regular lattices and their
thermodynamic functions in the complex-temperature
plane \cite{matshr}, which give clues on how to improve the
convergence behaviour of approximation methods applied
to high- and low-temperature expansions.

Going beyond regular lattices, and
staying within the same philosophy, the use of {\it random lattices}
has been advocated \cite{christ}, in the hope that the absence of
discrete lattice symmetries and therefore of distinguished
lattice directions may accelerate the restoration of continuous
rotational and translational invariances, and thus the approach
to the continuum limit. However, we are not aware of any applications in
lattice field theory where this would have led to a practical or conceptual
breakthrough.

Taking yet a further step toward randomizing the spaces
underlying the statistical models, one may consider an {\it additional
averaging over random lattices}. This is inspired by quantum
gravity, where the ``path integral" (a nonperturbative quantum
superposition of all spacetime geometries, which is central
to the quantum dynamics) can be defined via a statistical,
weighted sum over triangulated random geometries. % \cite{}.
The original approach of
{\it (Euclidean) Dynamical Triangulations} (EDT -- see \cite{david,livrev} for reviews)
turns out to be unsuitable for our purposes,
because the contributing triangulated lattices
are highly curved, with an effectively fractal structure, for any dimension
$d\geq 2$.
Their geometry is so radically different from the usual flat lattices
that it alters the universal behaviour of matter systems defined
on them, as has been well documented
for two-dimensional spin systems (see \cite{spin} and references therein).

In fact, these geometries are so ``wild" that they are not even suited for
modelling the quantum behaviour of four-dimensional gravity.
However, a promising new avenue has opened up recently with the
advent of {\it Causal Dynamical Triangulations (CDT)}, which were exactly
invented to fix the extreme geometric degeneracies of the previous,
Euclidean approach (see \cite{cdtrev} for a review).
One still works with an ensemble of lattices with
large local curvature fluctuations, but one where a partial order
has been imposed in one of the lattice directions (``time").
In ``pure" gravity (i.e. without matter coupling)
this is sufficient to produce geometries whose effective
(or Hausdorff) large-scale
dimension $d_H$ -- in the sense of ensemble
averages -- equals the dimension $d$ of their microscopic
triangular building blocks, for $d=2$ \cite{al}, $d=3$ \cite{3dcdt} and
$d=4$ \cite{ajl-prl}, which was {\it not} the case for the
corresponding Euclidean models, and is an indication that the geometries
are much better behaved.

\subsection*{Are causal dynamical triangulations the ``better lattices"?}

We will in what follows concentrate on the model of causal
dynamical triangulations in dimension $d=2$,
where the partition function of
pure quantum gravity has been computed exactly
\cite{al}. There is no known exact solution in the presence of
Ising spins, but CDT coupled to one \cite{aal1} and eight
\cite{aal2,aal3} Ising models has been studied using Monte Carlo
simulations.\footnote{Regarding the lattice fluctuations as
a type of disorder, the CDT-plus-matter models are examples
of models with so-called {\it annealed} disorder, i.e. there is
a genuine backreaction of the matter on the geometry.}
A rather surprising outcome\footnote{especially for those closely
familiar with the analogous Euclidean EDT results \cite{kaz}}
of this analysis was that the
behaviour of the matter is very robust:
to high accuracy, the critical matter exponents coincide
in both cases with
those of the Onsager solution, despite large fluctuations of
the underlying lattice geometry, and -- in the case of the eight
Ising copies -- a shift in the geometry's Hausdorff dimension $d_H$
from 2 to 3, due to the presence of the matter. This provides
strong evidence (though not a proof) that the universal
properties of any spin or matter system coupled to
two-dimensional causal dynamical triangulations will be
identical to those on a fixed, flat two-dimensional
lattice. Since in terms of the degree of geometric disorder our model
presumably lies in between the highly disordered EDT models
and the more mildly disordered random Voronoi-Delaunay
lattices, this would also lend credence to findings that the
three-state Potts model on the latter exhibits a behaviour
unchanged from the flat-lattice case \cite{potts}.

Assuming that the universal flat-space properties of
the matter systems do remain unaffected when coupled to
causal dynamical triangulations, we want to advocate
the CDT ensemble of triangulated geometries as a
``background lattice" for studying their continuum behaviour.
In view of the fact that these geometries are presumably
the ``maximally disordered" lattices with this property,
we will investigate the hypothesis that this will improve maximally the
approach to the scaling limit.

One way to test this hypothesis would be a systematic
numerical Monte Carlo study, comparing CDT-coupled
matter models to the same models on fixed regular or
random lattices. However, for the time being we will take
a different route, and try to understand
how far the issue can be pushed (semi-)analytically, by
exploiting another remarkable property of these models.
Namely, despite the absence of a {\it single} lattice
of known geometry, series expansions for thermodynamic
functions {\it can} be evaluated diagrammatically on the CDT
ensemble, both at
high and low temperature.
We have solved the associated combinatorial
problem by devising algorithms for counting graphs,
both for spin variables at the vertices
and at the centres of the triangles. Since this
constitutes to our knowledge a qualitatively new way of
setting up an asymptotic power series expansion,
this is of interest in its own right. It may also yield information
on the analytic structure of the magnetic susceptibility, say,
in a statistical model that is still relatively unexplored. This
in turn could provide a new reference point for the study of spin
models on a variety of regular and random lattices (see
\cite{janke} for a review of some current problems for the latter).

After recalling some necessary ingredients from the
original CDT model \cite{al,aal1}, we will give a sketch
of the combinatorial algorithms (details of which can be
found in our forthcoming publication \cite{bl}), and then
present our results. For Ising spins located at vertices,
we have evaluated the high-$T$ series expansion for
the magnetic susceptibility to order six.
Evaluating the series by straightforward ratio
method gives a remarkably good result for the critical
susceptibility exponent $\gamma$, compared with
that of the corresponding fixed, regular lattice.
We have also computed the same
expansion for the Ising model on dual CDT lattices, up
to order 12. As expected, the data are less good, but
also demonstrate clearly the simplified
singularity structure of the susceptibility function.

We are very encouraged by these results. Though not
conclusive with regard to our hypothesis, they provide sufficient
evidence for the viability of the method, and
warrant a more extensive investigation of the critical
properties of the underlying triangulated spin model. In view of
the computational effort involved, the diagram counting can then
no longer be done by hand, but will at least in part have to be
computerized.
Given the random nature of the graphs, this certainly presents a
non-trivial task, but one that we believe is feasible.

\subsection*{Method}

\begin{figure}[t]
%\vspace{-3cm}
\centerline{\scalebox{0.5}{\rotatebox{0}{\includegraphics{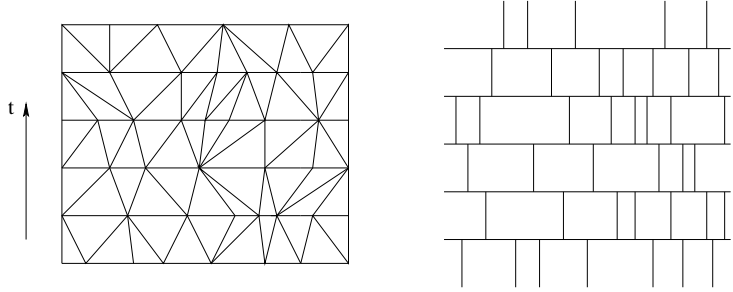}}}}
%\vspace{-1cm}
\caption[]{{\small A piece with five time steps of a triangulation contributing
to the CDT ensemble of geometries (left), and its corresponding dual lattice
(right). }}
\label{cdt-vs-dual}
\end{figure}

In the approach of causal dynamical triangulations, the gravitational path integral over
two-dimensional causal spacetimes is represented by
the discrete sum
\begin{equation}
G(N,t)=\sum_{{\rm T}\in {\cal T}_{N,t}} 1
\label{pfgrav}
\end{equation}
over triangulations
${\rm T}\in {\cal T}_{N,t}$ of a fixed number
$N$ of triangles and $t$ of time steps.
A piece of such
a spacetime is depicted in Fig.\,\ref{cdt-vs-dual} (left).
Coupling the pure gravity model to Ising spins $\sigma_i=\pm 1$
in a magnetic field $H$,
one obtains the partition function
\begin{equation}
\tilde Z(N,t,\beta,H)=\sum_{{\rm T}\in {\cal T}_{N,t}} \sum_{\{\sigma_i({\rm T})\} }
\exp\left(\beta \sum_{\langle ij\rangle\in {\rm T}}\sigma_i\sigma_j +H\sum_i
\sigma_i\right),
\label{pfising}
\end{equation}
with $\beta =J/k_BT$, and $\langle ij\rangle$ denoting a pair of
nearest neighbours, corresponding to an edge of the triangulation
or one of the trivalent graph dual to the triangulation
(Fig.\,\ref{cdt-vs-dual} right), depending
on whether the spins reside on the vertices or at the centres of the
triangles. The ferromagnetic Ising coupling has been set to $J=1$.

Turning now our interest to the high-$T$ expansion, we reexpress
(\ref{pfising}) as power series in the customary variables
$u=\tanh\beta$ and $\tau=\tanh H$, obtaining %\cite{aal1,bl}
\begin{equation}
\tilde Z(N,t,\beta,H)=
(\cosh\beta)^{3N/2} (2\cosh H)^{N/2}
\sum_{{\rm T}\in {\cal T}_{N,t}}
\left( 1+ \sum_{s=0}^\infty \sum_{n=1}^\infty
D^{(N)}_{n,2s}({\rm T})\ u^n\tau^{2s} \right),
\label{expand}
\end{equation}
where $D^{(N)}_{n,2s}({\rm T})$
counts graphs with $n$ edges and $2s$
odd vertices on the triangulation $\rm T$. It follows that
the magnetic susceptibility at vanishing $H$ is (up to a constant
term) given by \cite{aal1,bl}
\begin{equation}
\chi (u)=\frac{1}{N}\frac{\partial^2\ln \tilde Z}{\partial H^2}|_{H=0}=
\frac{2}{N} (\sum_{{\rm T}\in {\cal T}_{N,t}} 1)^{-1}
\sum_{n=1}^\infty \left( \sum_{{\rm T}\in {\cal T}_{N,t}}
\tilde D^{(N)}_{n,2}({\rm T})\right) u^n,
\label{sus}
\end{equation}
where the tilde on $\tilde D$ indicates that we count only
graph contributions of order $N^m$ with $m=1$. The crucial observation
of \cite{aal1} was that the expectation values per lattice vertex
\begin{equation}
\langle \tilde D_{n,2} \rangle :=
\frac{1}{N}
(\sum_{{\rm T}\in {\cal T}_{N,t}} 1)^{-1}
\sum_{{\rm T}\in {\cal T}_{N,t}} \tilde D^{(N)}_{n,2}({\rm T})
\label{expect}
\end{equation}
can in the continuum limit $N\rightarrow\infty$ be computed
from the known probability distribution of time-like edges
(the non-horizontal edges of Fig.\,\ref{cdt-vs-dual}, left, emanating
from a given vertex forward and backward in time) for the
{\it pure} gravity theory (which is the distribution relevant at both
$\beta=0$ and $\beta=\infty$).
The characteristic features of this probability
distribution are that (i) for a given time step $[t,t+1]$, we have
\begin{equation}
p_k(v)=\frac{1}{2^k}
\label{prop}
\end{equation}
for the probability that $k\geq 1$ time-like links emanate from
a given vertex $v$ on the lower time-slice $t$ toward the ``future"
(increasing time), {\it independently} for all vertices on the
slice, and (ii) there are no correlations between time-like
edges from different time steps $[t,t+1]$. Note that the absence of
a similar structure of the local probabilities would
prevent the direct application of this method to the case of
Euclidean dynamical triangulations.
\begin{figure}[t]
%\vspace{-3cm}
\centerline{\scalebox{0.6}{\rotatebox{0}{\includegraphics{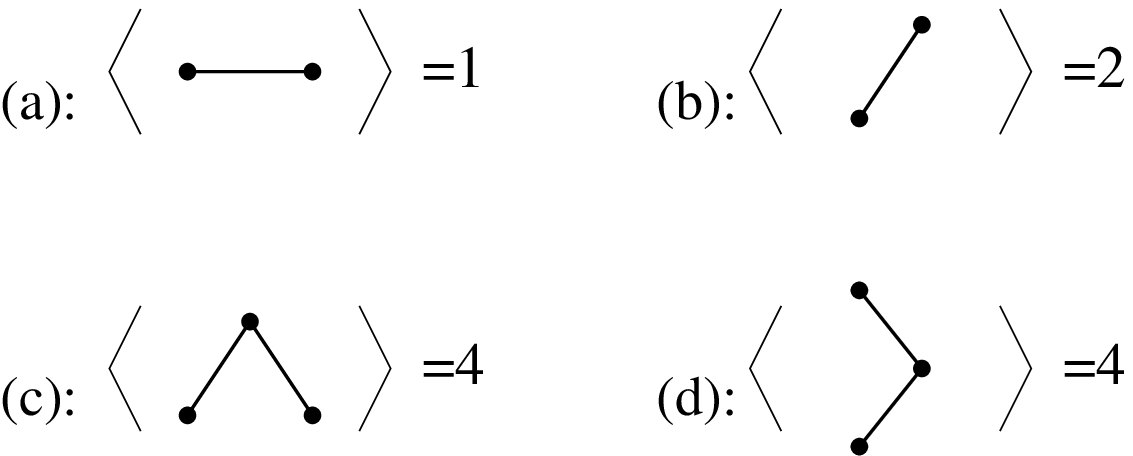}}}}
%\vspace{-1cm}
\caption[]{{\small Expectation values for the counting of some simple
two-odd susceptibility graphs on CDT lattices. }}
\label{examples}
\end{figure}

\subsection*{Results and discussion}

Some illustrative examples of expectation values for particular two-odd
graphs contributing to the calculation of the high-$T$ susceptibility
for Ising spins located at vertices are shown in Fig.\,\ref{examples}.
They already include appropriate multiplicities in case the
graph is not identical with its mirror image under left-right and/or
up-down reflection. The systematic algorithm we have devised
to calculate the contributions to (\ref{sus}) at order $n$ relies on
a ``strip decomposition" of a given graph into a sequence of
time strips $\Delta t=1$ \cite{bl}, thus exploiting the independence of
the probability distributions for distinct strips. As can be seen
from Table 1, the number of graphs rises quite rapidly with $n$.
To give an idea of the computational effort involved, there are
387 distinct open graphs which contribute at order 6 and whose
computation requires a careful use of the strip decomposition
and the associated algebra. Our results
confirm and extend those cited in \cite{aal1}.
\begin{center}
\vspace{.5cm}
\begin{tabular}{|r||r|r|r|r|r|r|}\hline
\multicolumn{7}{|c|}{{\rule[-3mm]{0mm}{8mm}
Table 1: Order-$n$ susceptibility graphs per vertex, CDT lattice.}} \\
\hline
\hline
{\rule[-3mm]{0mm}{8mm} $n$} & 1 & 2 & 3 & 4 & 5 & 6  \\
\hline\hline
{\rule[-3mm]{0mm}{8mm} $\langle \tilde D_{n,2} \rangle$}
& \hspace{.7cm} 3 & \hspace{.8cm}17 & \hspace{.8cm}87 &
\hspace{.3cm} 423$\frac{2}{3}$ & \hspace{.5cm}1995 & 9192$\frac{8}{27}$  \\
 \hline
%\caption{Susceptibility coefficients}
\end{tabular}
\vspace{.5cm}
\end{center}
To obtain more data on the behaviour of asymptotic power series in
the CDT framework, we next considered the high-$T$ expansion
of the susceptibility on the ensemble of random lattices {\it dual} to CDT,
corresponding to Ising spins placed at the centres of the triangles of
the original lattices. By virtue of the fact that the dual graphs are
trivalent, one can -- despite their randomness -- apply a powerful
counting theorem \cite{count}, which relates contributions of different
orders $n$ recursively. We refer to \cite{bl} for details of the counting
algorithm. Another reason for why in this case
we could push the evaluation to
order 12 is that there are considerably fewer distinct graphs on the
dual trivalent random lattice than there are on the original lattice,
where {\it any} number of edges (larger than 3) can meet at a vertex.
The results of the counting (per vertex of the {\it original}
lattice) are summarized in Table 2.
\begin{center}
\vspace{.5cm}
\begin{tabular}{|r|r|r|r|r|r|r|r|r|r|r|r|r|}\hline
\multicolumn{13}{|c|}{{\rule[-3mm]{0mm}{8mm}
Table 2: Order-$n$ susceptibility graphs per vertex, dual CDT lattice.}} \\
\hline
\hline
{\rule[-3mm]{0mm}{8mm} $n$} & 1 & 2 & 3 & 4 & 5 & 6 & 7
& 8 & 9 & 10 & 11 & 12\\
\hline\hline
{\rule[-3mm]{0mm}{8mm} $\langle \tilde D_{n,2} \rangle$} & 3 & 6 & 12 & 23
& 42$\frac{3}{4}$ & 78$\frac{1}{2}$ & 142$\frac{3}{4}$
 & 258 & 461$\frac{13}{16}$ & 820$\frac{1}{8}$ & 1446$\frac{13}{32}$
& 2532$\frac{11}{16}$\\
 \hline
%\caption{Susceptibility coefficients}
\end{tabular}
\vspace{.5cm}
\end{center}

\begin{figure}[ht]
%\vspace{-3cm}
\centerline{\scalebox{1.0}{\rotatebox{0}{\includegraphics{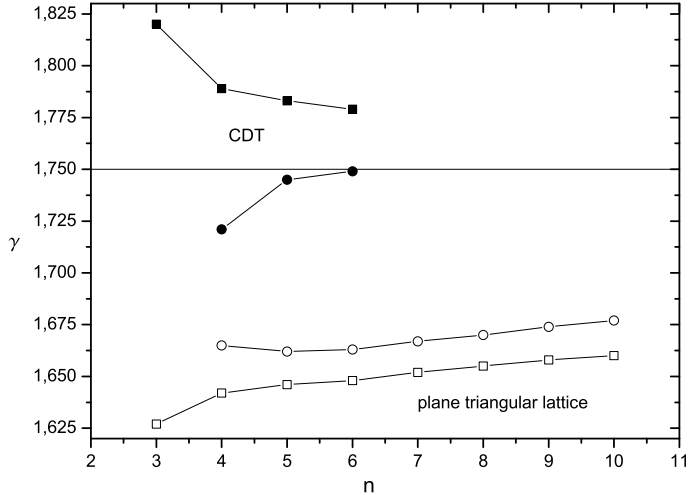}}}}
%\vspace{-1cm}
\caption[]{{\small The critical susceptibility exponent $\gamma$, obtained
from an order-6 high-$T$ expansion on CDT lattices (solid dots). The
shorter curve has the lowest-order ratio omitted. The analogous curves
for the regular triangular lattice (up to order 10) are given for comparison.
}}
\label{gamma}
\end{figure}

In order to get an idea of the quality of these data, we have compared
them with the
corresponding regular triangular and honeycomb lattices for which similar
expansions (to order 16 and 32 respectively) are available \cite{regular}.
In line with our conjecture that the singularity structure
with the use of CDT lattices should be maximally simplified,
we make the simplest possible ansatz for $\chi$, namely,
\begin{equation}
\chi (u)\sim A(u) \Bigl( 1-\frac{u}{u_c}\Bigr)^{-\gamma}+B(u)
\label{ratio}
\end{equation}
near the critical point $u_c$, with analytical functions $A$ and $B$.
Using the ratio method, which is known to work best
in the absence of interfering unphysical singularities
\cite{baker}, we have fitted the ratios of subsequent
susceptibility coefficients to
\begin{equation}
r_n=\frac{\langle \tilde D_{n,2} \rangle}
{\langle \tilde D_{n-1,2} \rangle}=\frac{1}{u_c}\left(1+
\frac{\gamma -1}{n}\right).
\label{rndefine}
\end{equation}
By plotting the $r_n$ linearly against $1/n$, we have
extracted simultaneous estimates for the critical point
$u_c$ and the critical exponent $\gamma$, the latter of
which is illustrated in Fig.\,\ref{gamma} as function of
$n_{\rm max}$. The convergence to the known true value
$\gamma =7/4$ is rather striking, especially when one
removes the lowest-order ratio (which is maximally affected
by omitting higher-order terms in $1/n$).
\begin{figure}[t]
%\vspace{-3cm}
\centerline{\scalebox{1.0}{\rotatebox{0}{\includegraphics{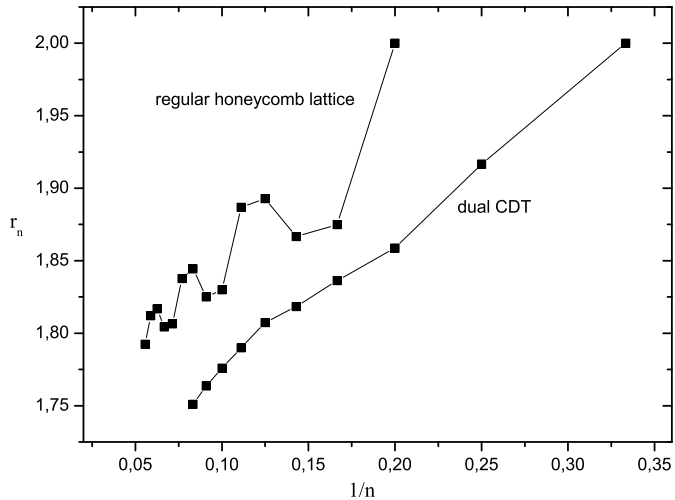}}}}
%\vspace{-1cm}
\caption[]{{\small A comparative plot of the ratios $r_n$ of the regular
honeycomb lattice versus those of
the dual CDT model, as function of $1/n$.
}}
\label{HCvsDual}
\end{figure}
Further evidence of how the use of dynamical
lattices seems to simplify the singularity
structure is the model with Ising spins at the
vertices of the dual, trivalent CDT lattices.
A priori one would not expect the data to be equally
good, because triangles (or, equivalently, lattices of
high coordination number) tend to explore the configuration
space more effectively. Nevertheless, a plot of the
ratios $r_n$ as function of $1/n$ (Fig.\,\ref{HCvsDual})
reveals a {\it much} smoother curve than that for
the corresponding regular, trivalent honeycomb lattice,
justifying again the use of the straightforward
ratio method (see \cite{bl} for details). This is likely due to the absence
of an antiferromagnetic phase of
the model\footnote{A similar effect was already noted in an
EDT-Ising model in \cite{fat}.}, implying the absence of the
interfering equidistant singularity on the negative
real axis of the complex $u$-plane of the regular lattice, as well as
to the possible absence of the
equidistant ``segments" along the imaginary axis \cite{matshr},
an issue that deserves further investigation.

Note that
we are {\it not} advocating here the use of the ratio method
for analyzing the susceptibility coefficients for
the {\it regular} honeycomb lattice; if one wants to stick to
a regular lattice, there clearly
are better ways of trying to take into account the influence
of the unphysical singularities on the negative
real axis and in the complex plane evident in
Fig.\,\ref{HCvsDual} (see, for example, \cite{regular,guttmann}).
Rather, we are advocating the use of dynamical CDT lattices
instead of fixed, regular lattices as a general method for determining the
critical behaviour of spin and matter systems, for cases
which are not
already known by other methods.
The strategy we suggest here is to first test our main conjecture more
thoroughly for other two-dimensional spin systems (e.g. higher-$q$
Potts and $O(n)$-models) for which our exact counting
method can be applied, and then move on to higher-dimensional
CDT-matter models, about whose analytic structure and exact
critical properties much less is known.
By eliminating
maximally the influence of spurious singularities present for
regular lattices, they will
(hopefully) lend themselves to straightforward approximation
methods without the need for any {\it lattice-specific}
subtraction schemes.

The results obtained so far for the well-known ``test case" of
the Ising susceptibility are interesting and encouraging.
We believe it is not accidental that the high-$T$ expansions
give good results with the simple ratio method and for
relatively few terms in the expansion, but we ultimately
do not know why this is so without a more detailed
understanding of the critical structure of these coupled
CDT-spin models. We have also computed part of the low-$T$
expansion \cite{bl} (which in the case of regular lattices
is notorious for the interference of unphysical
complex-$T$ singularities \cite{low,matshr}), but the series
we obtain are simply too short to draw any conclusions
one way or the other. The same seems to be true when
we try to use more elaborate methods to extract information
about the critical properties from these finite series, such
as the Dlog Pad\'e and differential approximants \cite{bl}.
With the counting algorithms
for two-dimensional CDT lattices in place,
the method now needs to be computerized, so that longer
expansions can be obtained, evaluated, and compared to
known results on regular lattices.
Besides paving the way for a possible application to higher-dimensional
matter systems, the two-dimensional analysis is likely to shed further
light on recent attempts to formulate quantitative criteria for when
geometric randomness of an underlying lattice is relevant to the
critical behaviour of a matter system defined on it.
For example, if the three-state Potts model coupled to CDT exhibited
the critical exponents of the flat-lattice model, and if the local
fluctuations of geometry could be shown to be sufficiently uncorrelated
(both of which we expect to be true), it would provide a
counterexample to the so-called Harris-Luck criterion \cite{harrisluck,janke}.
Complementary to the use of
approximation methods based on series expansions, one can also
explore
different matter types and higher-dimensional models purely
numerically. A study of the physically most interesting case of
four dimensions is currently under way \cite{ajlmatter}.

%\subsection*{Conclusions}

\vspace{.7cm}

\noindent {\bf Acknowledgement.} Both authors are partially supported
through the European Network on Random Geometry
ENRAGE, contract MRTN-CT-2004-005616. R.L. acknowledges
support by the Netherlands Organisation for Scientific Research
(NWO) under their VICI program.

\end{document}